\documentclass[12pt]{article}
\usepackage{amssymb}
\usepackage{amsmath}
\usepackage{epsf}
\usepackage{epsfig}
\usepackage{psfrag}
\usepackage{rotating}
\sffamily
\allowdisplaybreaks[1]

\newcommand{\lsim}{
\mathrel{\hbox{\rlap{\hbox{\lower4pt\hbox{$\sim$}}}\hbox{$<$}}}}

\newcommand{\gsim}{
\mathrel{\hbox{\rlap{\hbox{\lower4pt\hbox{$\sim$}}}\hbox{$>$}}}}

\newcommand{\vcb}{|V_{cb}|}
\newcommand{\vtd}{|V_{td}|}

\newcommand{\vts}{|V_{ts}|}
\newcommand{\vus}{|V_{us}|}

\newcommand{\gev}{\, {\rm GeV}}
\newcommand{\mev}{\, {\rm MeV}}

\newcommand{\be}{\begin{equation}}
\newcommand{\ee}{\end{equation}}
\newcommand{\bi}{\begin{itemize}}
\newcommand{\ei}{\end{itemize}}
\newcommand{\ord}{{\cal O}}

\def\kpn{K^+\rightarrow\pi^+\nu\bar\nu}
\def\klpn{K_{\rm L}\rightarrow\pi^0\nu\bar\nu}

\begin{document}
\begin{titlepage}
\vspace*{-0.5truecm}

\begin{flushright}
TUM-HEP-573/05\\
hep-ph/0501230
\end{flushright}

\vspace*{0.3truecm}

\begin{center}
\boldmath

{\Large{\bf Non-Decoupling Effects of the Heavy $T$ 
\vspace{0.3truecm}

in the $B^0_{d,s}-\bar B^0_{d,s}$ Mixing and 
\vspace{0.3truecm}

Rare $K$ and $B$ Decays 
}}

\unboldmath
\end{center}

\vspace{0.4truecm}

\begin{center}
{\large\bf Andrzej J. Buras, Anton Poschenrieder
and Selma Uhlig} 
\vspace{0.4truecm}

{\sl Physik Department, Technische Universit\"at M\"unchen,
D-85748 Garching, Germany}

\vspace{0.2truecm}

\end{center}

\vspace{0.6cm}
\begin{abstract}
\vspace{0.2cm}\noindent

 We point out that in the case of a heavy top quark $T$, present in the
Littlest Higgs model (LH), and the $t$-$T$ mixing parameter $x_{L} \geq 0.9$ 
the contribution to 
$B^0_{d,s}-\bar B^0_{d,s}$ mixing from box diagrams with {\it two} 
$T$ exchanges cannot be neglected. 
Although formally $\ord(v^4/f^4)$ with $v=246\gev$ and 
$f> 1~{\rm TeV}$, this contribution 
increases linearly with
$x_T=m_T^2/M^2_W$ and with $x_T=\ord(f^2/v^2)$  constitutes effectively
an $\ord(v^2/f^2)$ correction. 
For $x_L\approx 1$, this contribution 
turns out to be  more important than the genuine
$\ord(v^2/f^2)$ corrections. In particular it is larger than 
the recently calculated $\ord(v^2/f^2)$ 
contribution of box diagrams with a {\it single} 
$T$ exchange that increases only logarithmically with $x_T$. 
For $x_L=0.95$ and $f/v=5,10,15$, the short distance function $S$ 
governing the $B^0_{d,s}-\bar B^0_{d,s}$ mixing mass differences 
$\Delta M_{d,s}$ receives  $56\%$, $15\%$ and $7\%$ enhancements relative 
to its Standard Model (SM) value,  
implying a suppression of the CKM element $\vtd$ and 
an enhancement of $\Delta M_s$.  
The short distance functions $X$ and $Y$, relevant for rare $K$ and $B$ 
decays, increase only logarithmically with $x_T$. With the suppressed $\vtd$, 
$K\to\pi \nu\bar\nu$ and $B_d\to\mu^+\mu^-$ decays are only insignificantly 
modified with respect to the SM, while the branching ratio 
$Br(B_s\to\mu^+\mu^-)$  receives  $66\%$, $19\%$ and $9\%$ enhancements
for $x_L=0.95$ and $f/v=5,10,15$, respectively. 
Similar enhancement is found for 
$Br(B_{s}\to\mu\bar\mu)/Br(B_{d}\to\mu\bar\mu)$.

\end{abstract}

\end{titlepage}

\thispagestyle{empty}
\vbox{}

\setcounter{page}{1}
\pagenumbering{roman}

\setcounter{page}{1}
\pagenumbering{arabic}
\section{Introduction}\label{sec:intro}
It is well known that in the Standard Model (SM), 
flavour changing neutral current 
processes (FCNC) such as 
$B^0_{d,s}-\bar B^0_{d,s}$ mixing, CP violation in $K\to\pi\pi$
 and rare $K$ and $B$ decays are dominated 
by the contributions of top quark exchanges in box and penguin diagrams 
\cite{IL,Schladming}. 
This dominance originates in the large mass $m_t$ of the top quark and in 
its non-decoupling from low energy observables due to the corresponding Yukawa 
coupling that is proportional to $m_t$. In the evaluation of box and
penguin diagrams in the Feynman-t'Hooft gauge this decoupling is realized
through the diagrams with internal fictitious Goldstone boson and top quark
exchanges. The couplings of Goldstone bosons to the top quark, being
proportional to $m_t$, remove the suppression of the diagrams in question 
due to top quark propagators so that at the end the box and penguin diagrams
increase with increasing $m_t$. In the unitary gauge, in which fictitious
Goldstone bosons are absent, this behaviour originates from the 
longitudinal $(k_\mu k_\mu/M^2_W)$ component of the $W^\pm$--propagators.

In particular, in the case of $B^0_{d,s}-\bar B^0_{d,s}$ mixing  
in the SM the relevant $m_t$ dependent Inami-Lim function 
$S(x_t)\equiv S_0(x_t)$ \cite{IL,BSS} has the following large $m_t$ behaviour 
\cite{Schladming}
\be\label{Sasym}
S(x_t) \to \frac{x_t}{4}, \qquad  x_t=\frac{m_t^2}{M_W^2}.
\ee 
Similarly the functions $X(x_t)$ and $Y(x_t)$, relevant for instance for 
$K\to\pi\nu\bar\nu$ and $B_{d,s}\to\mu^+\mu^-$, respectively, have the 
following large $m_t$ behaviour
\be\label{XYasym}
X(x_t) \to \frac{x_t}{8}, \qquad  Y(x_t) \to \frac{x_t}{8}.
\ee
Yet, with $x_t\approx 4.4$, these formulae are very poor approximations 
of the true values $S=2.42$, $X=1.54$ and $Y=0.99$. 
We will see below that in the 
case of the Littlest Higgs (LH) model, the value of the 
corresponding variable $x_T$ is 
at least $400$ and 
the asymptotic formulae presented below are excellent approximations of the 
exact expressions.

In the Little Higgs models \cite{LH1}-\cite{LH5}, that offer 
an attractive and a rather simple solution to  the gauge hierarchy problem,
there is a new very heavy top quark $T$.  In the LH model \cite{LH4} 
its mass is given by
\be \label{mT}
m_T=\frac{f}{v}\frac{m_t}{\sqrt{x_L (1-x_L)}}, 
\qquad
x_L=\frac{\lambda_1^2}{\lambda_1^2+\lambda_2^2}.
\ee
Here $\lambda_i$ parametrize the Yukawa interactions of the top quark and 
$v=246\gev$ is the vacuum expectation value of the SM Higgs 
doublet. The parameter $x_{L}$ enters the sine of the $t$-$T$ 
mixing which is  simply given by $x_{L} v/f$.
The new scale  $f> 1~{\rm TeV}$ is related to 
$\Lambda \sim 4 \pi f \sim \mathcal{O}\left(\textrm{10 TeV}\right)$ 
at which the gauge group of the LH model 
$\left[SU(2)_1\otimes U(1)_1\right]\otimes\left[SU(2)_2\otimes U(1)_2\right]$
 is broken
down to the SM gauge group. The SM results for various observables of 
interest receive $\ord(v^2/f^2)$ corrections that originate in new heavy
gauge boson and scalar exchanges and in particular in the diagrams with 
the heavy $T$.
 The constraints from various processes, 
in particular from electroweak precision observables and direct new particles
searches, have been extensively analyzed in \cite{Logan}-\cite{PHEN6}.

\begin{figure}[htb]
\vspace{0.10in}
\centerline{
\epsffile{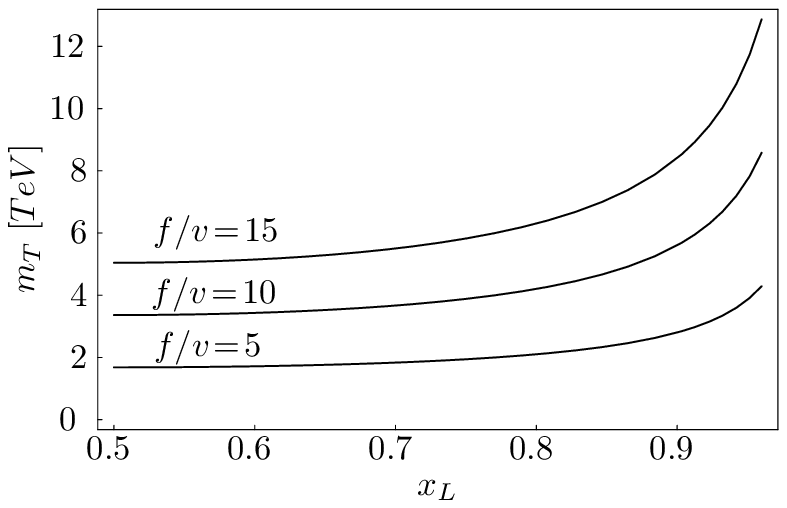}}
\vspace{0.01in}
\caption{The mass of $T$ as a function of $x_L$ for different values of $f/v$.
}
\label{fig:Tmass}
\end{figure}

As already discussed in \cite{Logan,BPU04}, the
parameter $x_L$ describes together with $v/f$ the size of the violation of
the three generation CKM unitarity and is also crucial for the gauge
interactions of the heavy $T$ quark with the ordinary down quarks. 
$x_L$ can in principle vary in the range $0<x_L<1$. For 
$x_L\approx 1$, the mass $m_T$ becomes large and its coupling
to the ordinary $W^\pm_L$ bosons and the down quarks, 
$W^\pm _L \bar T d_j$, being $\ord(x_L v/f)$, is only suppressed by $v/f$. 
In Fig.~\ref{fig:Tmass} we show the dependence of $m_T$ on $x_L$ for 
three values of $f/v$.

We are aware of the fact that with increasing $m_{T}$ also one-loop 
corrections to the SM Higgs mass increase. 
Typically for $m_{T} \geq 6~ \textrm{TeV}$ a fine-tuning of at least 
$1\%$ has to be made in order to keep $m_{H}$ below $200 \textrm{GeV}$
\cite{Csaki:2003si,Perelstein:2003wd}. 
As roughly $f/v \geq 8$ is required by electroweak precision studies 
\cite{Logan}-\cite{PHEN6},
the non-decoupling effects of $T$ considered here can be significant 
and simultaneously consistent with 
these constraints only in a narrow range of $f/v$. 
But these bounds are clearly  model 
dependent and  we will consider the range $5 \leq f/v \leq 15$ and 
$x_L\le 0.95$ 
for completeness.

As with $f/v\ge 5$ the $T$ quark is by an order of magnitude heavier than 
the ordinary top quark, it is of interest to ask what are its effects in 
the FCNC processes. In the present letter we summarize the main results 
of an analysis
of $T$ contributions to the function $S$ for $x_L$ close to unity, pointing
out that the existing $\ord(v^2/f^2)$ calculations of the function $S$ in 
the LH model \cite{BPU04,IND1} are inadequate for the description of this
parameter region. The $\ord(v^4/f^4)$ contributions involving
 the $T$ quark 
have to be taken into account as well and
in fact for $x_L>0.85$ these $\ord(v^4/f^4)$ corrections turn out to be 
as important as the genuine $\ord(v^2/f^2)$ corrections. For higher values 
of $x_L$ they become even dominant.
The details of these investigations will appear in an update of \cite{BPU04}.

On the other hand our analysis of the functions $X$ and $Y$ \cite{BPU05} 
shows that they increase with $x_T$ only as $\log x_T$ and that an 
$\ord(v^2/f^2)$ analysis is sufficient for the study of the large $x_T$ 
behaviour. Below we will summarize the main results of this analysis. 
The details will be presented in \cite{BPU05}.

\boldmath
\section{The Function $S$ and $\Delta M_{d,s}$ in the LH Model}
\unboldmath
\setcounter{equation}{0}
In \cite{BPU04} we have calculated the $\ord(v^2/f^2)$ contributions to the 
function $S$ in the LH model. Our results have been recently confirmed in 
\cite{IND1}. The
$\ord(v^2/f^2)$ correction $\Delta S$ can be written as follows
\be\label{Sold}
\Delta S=(\Delta S)_T+(\Delta S)_{W^\pm_H}+(\Delta S)_{\rm Rest}
\ee
with the three contributions representing the $T$, $W^\pm_H$ (new heavy
charged gauge bosons) and the remaining
contributions that result from $\ord(v^2/f^2)$ corrections to the vertices
involving only  SM particles. 

As demonstrated in \cite{BPU04} the three contributions in question  
are given to a very good approximation as follows ($W_L^\pm\equiv W^\pm$)
\be\label{DTS}
(\Delta S)_T = \left[\frac{1}{2}\frac{v^2}{f^2} x_L^2\right] x_t (\log x_T
-1.57), \qquad x_T=\frac{m_T^2}{M_{W_L^\pm}^2}~,  
\ee
\be\label{DWS}
(\Delta S)_{W^\pm_H}= 2 \frac{c^2}{s^2}\frac{m_t^2}{M^2_{W^\pm_H}}=2 
\frac{v^2}{f^2} c^4 x_t~,
\ee
\be\label{DRest}
(\Delta S)_{\rm Rest} = - 2 \frac{v^2}{f^2} c^4 S_{\rm SM}(x_t),
\ee
where $x_t$ has been defined in (\ref{Sasym}), $S(x_t)_{\rm SM}=S_0(x_t)$ in 
\cite{BPU04} and the numerical factor 
in (\ref{DTS}) corresponds to $m_t(m_t)=168.1\gev$.  The mixing 
parameters $s$ and $c$ \cite{Logan} are related through $s=\sqrt{1-c^2}$ 
with $0<s<1$.

For $ x_L\le 0.8$ considered in \cite{BPU04,IND1} and $s<0.4$ the most
important contribution turns out to be $(\Delta S)_{W^\pm_H}$. 
However, for higher 
values of $s$ and $x_L> 0.7$, the contribution $(\Delta S)_T$ becomes more 
important.

\begin{figure}[htb]
\vspace{0.10in}
\centerline{
\epsffile{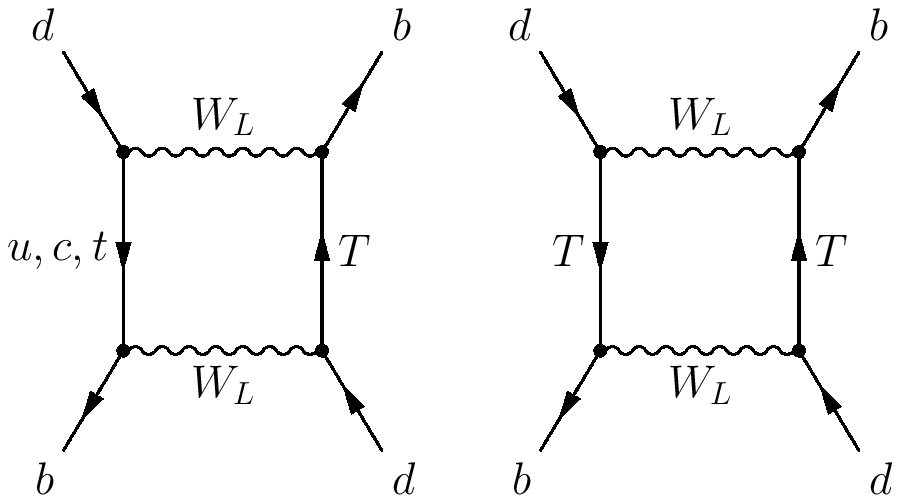}}
\vspace{0.01in}
\caption{ Single and double heavy Top  contributions 
to the function $S$  in
  the LH model at $\ord (v^2/f^2)$ and $\ord (v^4/f^4)$, respectively.}
\label{fig:boxTT}
\end{figure}

We observe that all three contributions have a characteristic linear
behaviour in $x_t$ that signals the non-decoupling of the ordinary top quark.
However, the corresponding non-decoupling of $T$ is only logarithmic. 
This is related 
to the fact that with the  $W^\pm_L \bar T d_j$ coupling being $\ord(v/f)$ 
only box diagrams with a single $T$ exchange (see Fig.~\ref{fig:boxTT}) 
contribute at 
$\ord(v^2/f^2)$. Similarly to the SM box diagrams with a single $t$ 
exchange, that increase as $\log x_t$, the $T$ contribution in the 
LH model increases only as $\log x_T$.

Here we would like to point out that for $x_L \geq 0.85$ , the 
term $(\Delta S)_T$ in (\ref{DTS}) does not give a proper
description of the non-decoupling of $T$. Indeed in this case also the
box diagram with two $T$ exchanges given in Fig.~\ref{fig:boxTT}
 has to be considered. 
Although formally $\ord(v^4/f^4)$, this contribution increases linearly with
$x_T$ and with $x_T=\ord(f^2/v^2)$  constitutes effectively
an $\ord(v^2/f^2)$ contribution.

Calculating the diagram with two $T$ exchanges in Fig.~\ref{fig:boxTT} 
 and adding those 
$\ord(v^4/f^4)$ corrections from
 box diagrams 
with $t$, $T$ and $u$ quark exchanges \cite{BPU04}, 
that have to be taken into account in
order to remove the divergences characteristic for a unitary gauge calculation
and for the GIM mechanism \cite{GIM} to become effective, we find 
\be\label{DTTS}
(\Delta S)_{TT}\approx\frac{v^4}{f^4} x_L^4 \frac{x_T}{4}
=
\frac{v^2}{f^2} \frac{x_L^3}{1-x_L} \frac{x_t}{4}~.
\ee
Formula (\ref{DTTS})  represents for $x_L>0.85$ and $f/v\ge 5$
 the exact expression given in \cite{BPU04} to within $3\%$ and becomes
rather accurate for $x_L>0.90$ and $f/v\ge 10.$

In fact the result in (\ref{DTTS}) can easily be understood.
$(\Delta S)_{TT}$ 
has a GIM structure \cite{BPU04}
\be\label{GIMLH}
(\Delta S)_{TT}=\frac{v^4}{f^4} x_L^4
\left[F(x_T, x_T; W_L)+F(x_t, x_t; W_L)
-{2} F(x_t, x_T; W_L)\right]
\ee
with the function $F(x_i, x_j; W_L)$ resulting up to an overall factor from
box diagram with two $W_L^{\pm}$ and two quarks ($i,j$)
exchanges. This GIM structure is identical to the one of $S$ in the SM 
that depends 
 on $x_t$ and $x_u$ and is given by
\be\label{S0}
S(x_t)=F(x_t, x_t; W_L)+F(x_u, x_u; W_L)-2 F(x_u, x_t; W_L).
\ee
For large $x_T$ 
it turns out to be a good approximation to 
evaluate $(\Delta S)_{TT}$ with
$x_t=0$. In this case (\ref{GIMLH})  reduces to
$S(x_t)$ in (\ref{S0}) with 
$x_t$ replaced by $x_T$ and $x_u$ by $x_t$. 
The factor ${x_t}/{4}$ in (\ref{Sasym}) 
is then replaced by $x_T/4$ as seen in (\ref{DTTS}).

\begin{figure}[htb]
\vspace{0.10in}
\centerline{
\epsffile{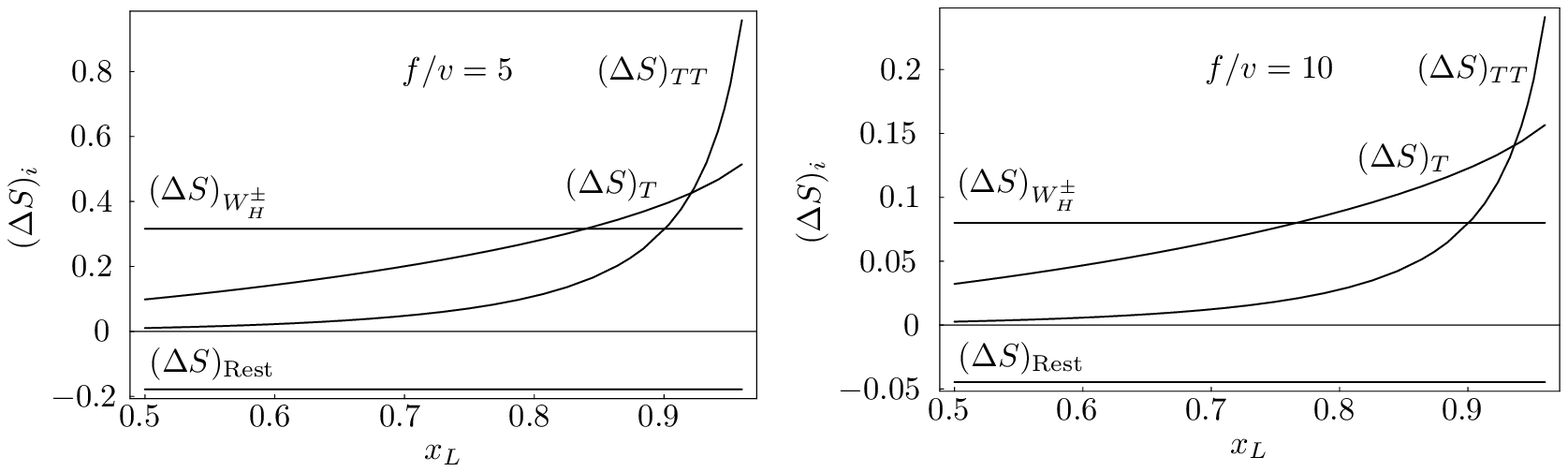}}
\vspace{0.01in}
\caption{The contributions $(\Delta S)_i$ as functions of $x_L$ for 
 $f/v=5$, $f/v=10$ and $s=\sqrt{1-c^2}=0.2$.
}
\label{fig:DSi}
\end{figure}

In Fig.~\ref{fig:DSi} 
we show the four contributions $(\Delta S)_i$ as functions of 
$x_L$ for $f/v=5$ and $f/v=10$ and $s=0.2$. We observe that
for $x_L<0.85$ the new contribution is  smaller than 
$(\Delta S)_T$ but for $x_L> 0.90$ it becomes a significant new effect
in $S$. 

\begin{figure}[htb]
\vspace{0.10in}
\centerline{
\epsffile{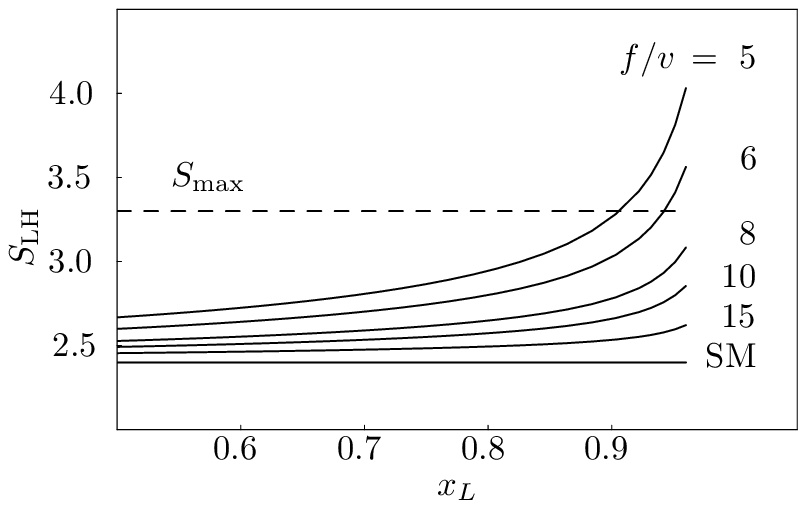}}
\vspace{0.01in}
\caption{$S_{\rm LH}$ as a function of $x_L$ for 
 different values of $f/v$, $s=0.2$. The dashed line 
is the bound in (\ref{Sbound}).
}
\label{fig:GE}
\end{figure}

A general upper bound on the function $S$ in  models with 
minimal flavour violation (MFV) can be obtained from the usual 
analysis of the unitarity triangle  \cite{BUPAST}. It is valid also in 
the LH model considered here. 
A recent update \cite{Stocchi} of this bound gives
\be\label{Sbound}
S\le 3.3, \qquad (95\%~~{\rm C.L.)}
\ee
to be compared with $S_{\rm SM}=2.42$ in the SM 
\cite{Schladming}. In Fig.~\ref{fig:GE} we plot 
\be\label{SLH}
S_{\rm LH}=S_{\rm SM}(x_t)+(\Delta S)_{TT}+
(\Delta S)_T+(\Delta S)_{W^\pm_H}+(\Delta S)_{\rm Rest} 
\ee
as a function of $x_L$ for different values of $f/v$ and $s=0.2$. 
We also show there the SM value and the upper bound in (\ref{Sbound}).
For a comparison we recall that from the studies of the $\rho$ 
parameter an upper bound on $x_L$ of $0.95$, almost independently of 
$f/v$, has been obtained \cite{PHEN3}. We observe that for $f/v=5$ a bound 
$x_L\le 0.90$ at $95\%\, {\rm C.L.}$ can be obtained. 
Weaker bounds are found for $5<f/v<10$ and in order to find 
a bound stronger than in \cite{PHEN3} for $f/v>10$, $S_{\rm max}$ should 
be within $10\%$ of the SM value. This would require significant 
reduction of the theoretical uncertainties in the analysis of the unitarity 
triangle.

With the precisely measured $\Delta M_d$ and $\vts\approx\vcb$, the 
observed enhancement of the function $S$ implies
\be\label{VTDMS}
\frac{(\vtd)_{\rm LH}}{(\vtd)_{\rm SM}}=
\sqrt{\frac{S_{\rm SM}}{S_{\rm LH}}}, \qquad
\frac{(\Delta M_s)_{\rm LH}}{(\Delta M_s)_{\rm SM}}=
\frac{S_{\rm LH}}{S_{\rm SM}},
\ee
that is the suppression of $\vtd$ and the enhancement of $\Delta M_s$.
As we will see below in the context of $K\to\pi\nu\bar\nu$ and 
$B_{d}\to\mu^+\mu^-$ decays, the suppression of $\vtd$ 
compensates to a large extent the enhancements of the functions $X$ and $Y$ 
in the relevant branching ratios. On the other hand, 
$\Delta M_s$ involving $\vts$ and being proportional to $S_{\rm LH}$, is for 
$x_L=0.95$ enhanced by 
$56\%$, $15\%$ and $7\%$  for $f/v=5,10,15$, respectively.

\boldmath
\section{Rare $K$ and $B$ Decays in the LH Model}
\begin{figure}[htb]
\vspace{0.10in}
\centerline{
\epsffile{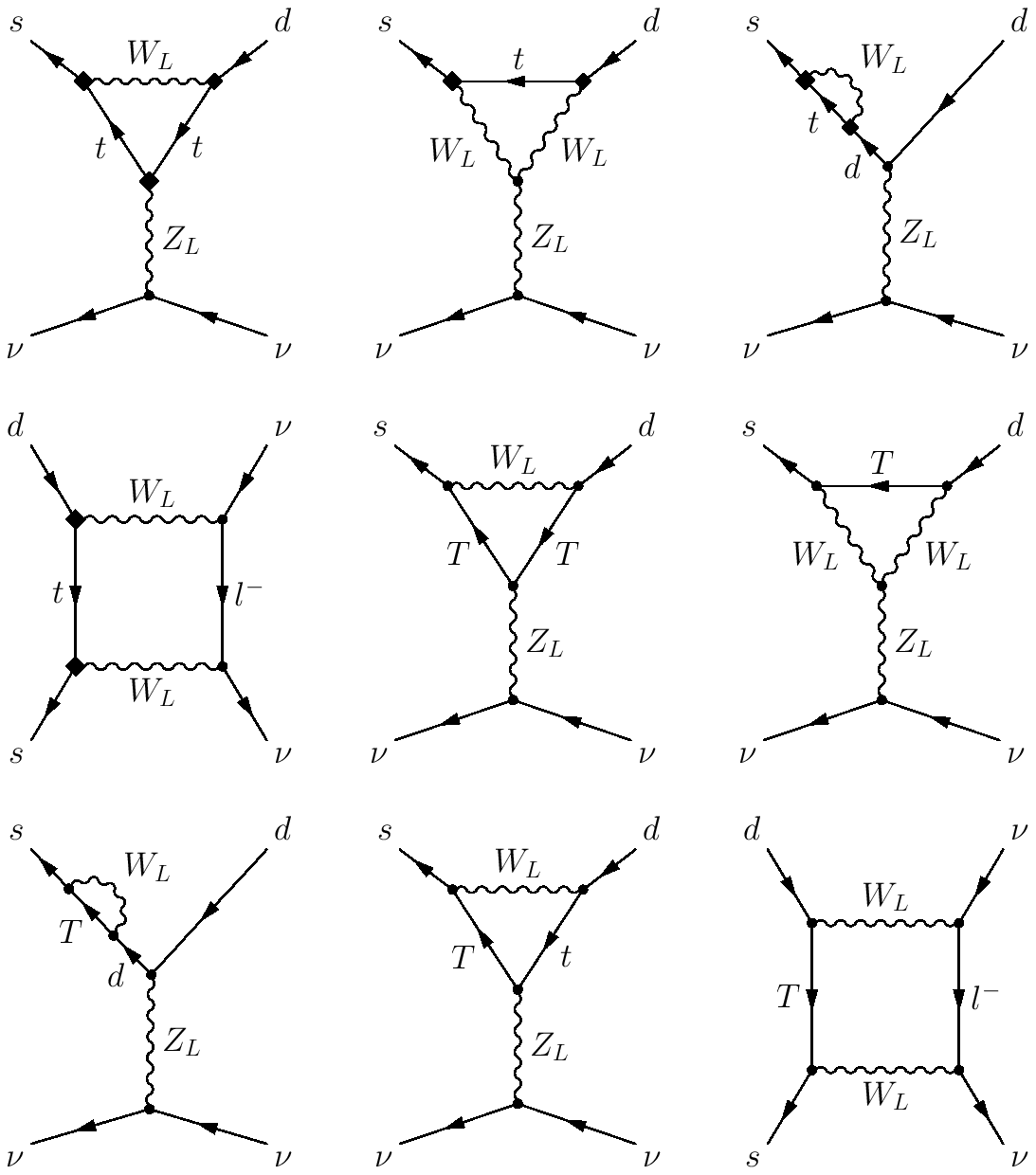}}
\vspace{0.01in}
\caption{ Top and heavy top quark contributions 
to the function $X$  in
  the LH model at $\ord (v^2/f^2)$ which are proportional to $x_L^2$.}
\label{fig:class3}
\end{figure}

\unboldmath
\setcounter{equation}{0}
The analysis of the functions $X$ and $Y$ is much more involved due 
to many box and penguin diagrams with $T$ and new heavy gauge boson
$Z^0_H$, $A^0_H$ and $W^\pm_H$ exchanges. The details of this analysis in 
the full space of the parameters involved will be presented in 
\cite{BPU05}. Here we present only the results for $x_L>0.7$, where 
the diagrams with the heavy $T$ and ordinary SM particles shown in 
Fig.~\ref{fig:class3} become dominant.

Due to a different topology of penguin diagrams and the fact that in 
the relevant box diagrams with $\nu\bar\nu$ and $\mu\bar\mu$ in the final
state only a single $T$ can be exchanged,  
the $\ord(v^2/f^2)$ diagrams in Fig.~\ref{fig:class3}
 give an adequate description
of the $x_L>0.7$ region. 
The diamonds in this figure 
indicate $\ord(x_L^2v^2/f^2)$ corrections to the SM vertices that are
explicitely given in \cite{Logan,BPU04}.
Many of the diagrams in this figure 
give in the unitary gauge contributions 
to $X$ and $Y$ that grow as $x_T$ and $x_T\log x_T$ but due to the GIM 
mechanism all these contributions cancel each other in the sum. In 
particular the penguin diagram involving two $T$ propagators, 
being proportional 
to $\sin^2\theta_w$, is canceled by other diagrams involving $\sin^2\theta_w$ 
so that $\sin^2\theta_w$ does not appear in the final result for $X$ and $Y$ 
as it should be.

We find then that to a very good approximation the corrections to $X$ and $Y$
in the LH model are given in the $x_L>0.7$ region as follows 
\be\label{XLH}
\Delta X=\left[\frac{v^2}{f^2} x_L^2\right]\left[(\frac{x_t}{4}+\frac{3}{8})
\log x_T-3.32\right], 
\ee
\be\label{YLH}
\Delta Y=\left[\frac{v^2}{f^2} x_L^2\right]\left[(\frac{x_t}{4}+\frac{3}{8})
\log x_T-3.53\right],
\ee
with the numerical factors 
corresponding to $m_t(m_t)=168.1\gev$. Exact formulae will be presented 
in \cite{BPU05}. 
A characteristic $x_t$-decoupling is observed with a logarithmic dependence 
on $x_T$. The combined dependence on $x_t$ and $x_T$ in this leading 
contribution makes it clear from which diagrams in Fig.~\ref{fig:class3}
 this leading 
contribution  comes from. 
This is the $Z^0$-penguin diagram with both $T$ and $t$ exchanges and 
the corresponding diagram with $t$ and $T$.

\begin{figure}[htb]
\vspace{0.10in}
\centerline{
\epsffile{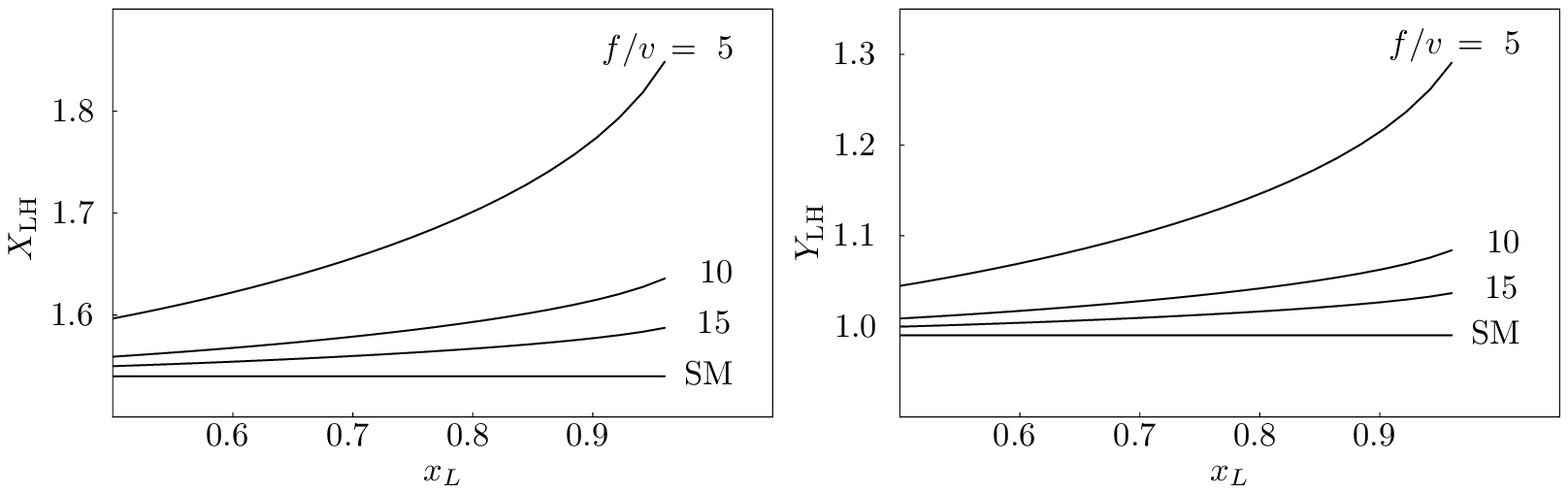}}
\vspace{0.01in}
\caption{$X_{\rm LH}$ and $Y_{\rm LH}$ as  functions of $x_L$ for 
different values of $f/v$.
}
\label{fig:XYLH}
\end{figure}

In Fig.~\ref{fig:XYLH} we show 
\be
X_{\rm LH}=X_{\rm SM}+\Delta X, \qquad 
Y_{\rm LH}=Y_{\rm SM}+\Delta Y
\ee
as functions
of $x_L$ for three values of $f/v$. We observe that due to the fact that 
$\Delta X\approx \Delta Y$ but $X_{\rm SM}\approx 1.6~Y_{\rm SM}$, the 
relative corrections to $Y$ are larger.

The branching ratios for $\klpn$ and $B_{s,d}\to\mu^+\mu^-$ are proportional 
to $X_{\rm LH}^2$ and $Y_{\rm LH}^2$, respectively.
The branching ratio for $\kpn$ grows slower with $X_{\rm LH}$ due to an
additive charm contribution that is essentially unaffected by the LH
contributions. Now, $\klpn$, $\kpn$ and $B_{d}\to\mu^+\mu^-$
involve the CKM element $V_{td}$, whose value is decreased in the LH model 
as discussed in the previous section. Consequently the enhancement of $X$ and
$Y$ 
in the LH model is compensated by the suppression of $V_{td}$ at the level of
the branching ratios. The situation is quite different in
$B_{s}\to\mu^+\mu^-$, where the relevant CKM element $\vts$ is 
approximately equal to 
$\vcb$ and the enhancement of $Y$ is fully visible in the branching ratio.
Explicitly we have
\be\label{RR1}
\frac{Br(\klpn)_{\rm LH}}{Br(\klpn)_{\rm SM}}
=\frac{S_{\rm SM}}{S_{\rm LH}}
\frac{X^2_{\rm LH}}{X^2_{\rm SM}},
\ee

\be\label{RR2}
\frac{Br(B_{d}\to\mu^+\mu^-)_{\rm LH}}{Br(B_{d}\to\mu^+\mu^-)_{\rm SM}}
=\frac{S_{\rm SM}}{S_{\rm LH}}
\frac{Y^2_{\rm LH}}{Y^2_{\rm SM}},
\ee

\be\label{RR3}
\frac{Br(B_{s}\to\mu^+\mu^-)_{\rm LH}}{Br(B_{s}\to\mu^+\mu^-)_{\rm SM}}
=\frac{Y^2_{\rm LH}}{Y^2_{\rm SM}}.
\ee

This pattern is clearly seen in Table~\ref{brtable2}. 
The values given there have been obtained by using the MFV formulae 
for branching ratios given in \cite{BSW} with 
$F_{B_d}=203\mev$, $F_{B_s}=238\mev$ and setting the three 
CKM parameters $\vus$, $\vcb$ and the angle $\beta$ in the 
unitarity triangle to
\be\label{CKMinput}
\vus=0.224, \qquad \vcb=0.0415, \qquad \beta=23.3^\circ.
\ee
The fourth parameter, the UT side $R_t$ is then calculated according to
\be\label{RtLH}
(R_t)_{\rm LH}=(R_t)_{\rm SM}\sqrt{\frac{S_{\rm SM}}{S_{\rm LH}}},
\qquad (R_t)_{\rm SM}=0.89 
\ee
with $(R_t)_{\rm SM}=0.89$ being the central value of the UT fit in 
\cite{BSU}.

We observe that $Br(\kpn)$ and $Br(\klpn)$ are very close to the SM value 
and with increasing $x_L$ they are even slightly suppressed. 
$Br(B_d\to \mu^+\mu^-)$ is slightly enhanced with respect to the SM, 
while the enhancement of $Br(B_s\to \mu^+\mu^-)$ for $f/v\le 10$ 
is significant.

\begin{table}[hbt]
\vspace{0.4cm}
\begin{center}
\caption[]{\small Branching ratios for rare decays in the LH model and the 
SM for $f/v=5$. In the last two rows the results for $f/v=10$ are given.
\label{brtable2}}
\begin{tabular}{|c||c|c|c|c|}
\hline
{$x_L$ } &  {$0.80$}&  {$0.90$} 
& {$0.95$} & SM
 \\ \hline
$Br(\kpn)\times 10^{11}$ &  $7.91$ & $ 7.78$ & $ 7.34$ &  $7.88$ 
\\ \hline
$Br(\klpn)\times 10^{11}$ &  $3.07$ & $ 3.00 $ & $ 2.76$ &  $3.05$ 
\\ \hline
$Br(B_d\to \mu^+\mu^-)\times 10^{10}$ &  $1.32$ & $ 1.34$ & $1.27$ &
$1.20$ \\ \hline
$Br(B_s\to \mu^+\mu^-)\times 10^{9}$ &  $5.17$ & $ 5.81 $ & $ 6.38$ &
$3.86$ \\ \hline
$R_{sd}$ &  $39.2$ & $ 43.4 $ & $ 50.4$ &
$32.2$ \\ \hline\hline
$Br(B_s\to \mu^+\mu^-)\times 10^{9}$ &  $4.27$ & $ 4.44 $ & $ 4.58$ &
$3.86$ \\ \hline
$R_{sd}$ &  $34.3$ & $ 35.4 $ & $ 37.2$ &
$32.2$ \\ \hline
\end{tabular}
\end{center}
\end{table}

The branching ratios $Br(B_{s,d}\to\mu\bar\mu)$ and $\Delta M_{s,d}$
are subject to uncertainties in $\vtd$ and the meson decay constants 
$F_{B_{s,d}}$. 
On the other hand the ratio
\be\label{RSD}
R_{sd}=\frac{Br(B_{s}\to\mu\bar\mu)}{Br(B_{d}\to\mu\bar\mu)}
\ee
is theoretically cleaner. We show this ratio in Table~\ref{brtable2}.

\section{Conclusions}\label{sec:Summ}
We have calculated the dominant $\ord(v^4/f^4)$ corrections to the function
$S$ in the LH model. Due to a large value of $m_T$, this contribution can
compete with the genuine $\ord(v^2/f^2)$ corrections. For
$x_L > 0.90$ it becomes a significant new contribution to the function $S$
in this model. As seen in (\ref{VTDMS}), the enhancement of $S$ implies 
the suppression of the value of $\vtd$ in the LH model and the enhancement 
of $\Delta M_s$ with the size of these effects depending sensitively on $x_L$ 
and $f/v$ (see Fig.~\ref{fig:GE}). With the improved precision of the
unitarity triangle analysis an upper bound on $x_L$, that is stronger than 
the one coming from the analysis of the $\rho$ parameter, could in 
principle be obtained.

The non-decoupling effects of $T$ in rare $K$ and $B$ decays discussed above
are weaker, with the functions $X$ and $Y$ growing only logarithmically 
with $x_T$.  We find that the branching ratios for 
$K\to\pi\nu\bar\nu$ and $B_{d}\to\mu^+\mu^-$ are only insignificantly
modified by the LH effects because the enhancements of $X$ and $Y$
 are compensated
by the decrease of $\vtd$. On the other hand 
$Br(B_{s}\to\mu^+\mu^-)$ can be enhanced up to $66\%$, $19\%$ and $9\%$ 
for $x_L=0.95$ and $f/v=5,~10,~15$, respectively.
This pattern is insignificantly modified through the effects of contributions
involving the new gauge bosons $Z^0_H$, $A^0_H$ and $W^\pm_H$ \cite{BPU05}.

The effects presented here are certainly of theoretical interest, although 
for $f/v\ge 8$ as required by other studies \cite{Logan}-\cite{PHEN6}, 
the corrections 
to the SM results for $S$, $X$ and $Y$ with $x_L$, even as high as 0.95, 
are only at most $15\%$ with the only relevant enhancements seen in $\Delta
M_s$ and $Br(B_{s}\to\mu^+\mu^-)$. On the 
other hand the effects found here could be larger in other Little 
Higgs models and our analysis of the function $S$ shows that 
$\ord(v^4/f^4)$ contributions
cannot be always neglected.

The non-decoupling of $T$ in the LH model has been already emphasized 
in \cite{IND2} in the context of $\klpn$. However, our result 
for this decay differs significantly from the large enhancement 
found by these authors. We will make a comparison with that paper
in \cite{BPU05}, where also other contributions to the rare decays 
in the LH model will be presented.

\noindent
{\bf Acknowledgements}\\
\noindent
We would like to thank  an unknown referee of \cite{BPU04} who
asked us to explain the result in (\ref{DTS}). This led us to 
reconsider our calculation of the function $S$.
We also thank Achille Stocchi for the bound in (\ref{Sbound}) and Andreas
Weiler for illuminating comments on our work.
The work presented here was supported in part by the German 
Bundesministerium f\"ur
Bildung und Forschung under the contract 05HT4WOA/3 and by the German-Israeli
Foundation under the contract G-698-22.7/2002. 




%
%
%
\end{document}